\documentclass{article}
\usepackage{graphicx}
\usepackage{dcolumn}
\usepackage{bm}

\def\be{\begin{equation}}
\def\ee{\end{equation}}
\def\ba{\begin{array}}
\def\ea{\end{array}}
\def\bea{\begin{eqnarray}}
\def\eea{\end{eqnarray}}

\begin{document}

\title{Study of fragmentation using clusterization algorithm with realistic binding energies}
\author{Yogesh K. Vermani, Jatinder K. Dhawan, Supriya Goyal and Rajeev K. Puri
\footnote{rkpuri@pu.ac.in} \\
Department of Physics, Panjab University, \\
Chandigarh-160014, India. \\
J. Aichelin \\
SUBATECH~-~IN2P3/CNRS - Ecole des Mines de Nantes \\
4, rue Alfred Kastler, F-44072 Nantes, Cedex 03, France}
\maketitle
\begin{abstract}

We here study fragmentation using \emph{simulated annealing
clusterization algorithm} (SACA) with binding energy at a
microscopic level. In an earlier version, a constant binding
energy (4 MeV/nucleon) was used. We improve this binding energy
criterion by calculating the binding energy of different clusters
using modified Bethe-Weizs\"{a}cker mass (BWM) formula. We also
compare our calculations with experimental data of ALADiN group.
Nearly no effect is visible of this modification.

\end{abstract}

\section{\label{intro}Introduction}
In the recent years, several theoretical attempts
\cite{barz,zheng,gosx,iwam,das} have been reported on spectator
matter fragmentation observed in relativistic heavy-ion (HI)
reactions using ALADiN set up \cite{hub,tsang,blaich,schut}. The
multifragmentation has been thought to be one of the important
phenomena for the understanding of phase transition in nuclei and
nuclear equation of state. The multiplicity of intermediate mass
fragments (IMFs) in central collisions is reported to first
increase with the beam energy with a peak at
$E\approx100~$MeV/nucleon \cite{tsang,peil} and then decline
afterwards indicating a complete disassembly of nuclear matter. At
relativistic energies, IMF emission becomes preferential only at
peripheral collisions \cite{hub,tsang,blaich,schut,ogi} where
system has relatively low excitation energy. The low energy
heavy-ion collisions are dominated by the phenomena such as the
deep-inelastic scattering and fusion-fission. The
fireball-spectator picture, however, emerges and dominates the
physics at relativistic energies where the formation of heavier
clusters is a rather unusual phenomenon. The most complete
experiments of ALADiN collaboration have shown that fragment
emission pattern remains almost unchanged above the incident
energy of 400 MeV/nucleon for a given projectile-target
combination \cite{schut}. This observation is also very often
termed as universality of the fragmentation emission and has been
discussed in the literature extensively
\cite{hub,blaich,schut,kreut}. In these experiments, the
correlation between IMF multiplicity and impact parameter `b'
suggests a picture of transition from evaporation to complete
disassembly with increasing violence of the collision
\cite{hub,blaich,ogi,kunde}. At higher incident energies, one also
expects complete disassembly or vaporization of the colliding
matter \cite{tsang,peil,peas}.

It has been reported earlier that fragment multiplicities
predicted by quantum molecular dynamics (QMD) model
\cite{aich,rkp,hart} coupled with conventional clustering
technique such as \emph{minimum spanning tree} (MST) algorithm are
significantly underestimated for larger values of impact
parameters \cite{gosx,blaich,tsang}. The choice of different
nuclear incompressibilities (\emph{i.e} equations of state) was
found to have only marginal influence on the predicted IMF
multiplicities and light charged particles yield \cite{blaich}.
About decade ago, Dorso \emph{et al.} \cite{dorso} advanced a new
algorithm in which fragments if already formed can be identified
earlier. The scope of this approach was limited to light systems
like Ca+Ca where only a few fragments are produced. The results
indicated a quite early formation. However, for the understanding
of multifragmentation, the multifragment events observed in the
collision of heavy systems have to be analyzed. Unfortunately, the
computing time for the algorithm employed \cite{dorso} increase by
roughly N!, where N is the number of nucleons in the system. Hence
a completely new numerical procedure was invented to extend the
approach to larger and more relevant systems \cite{saca}. Due to
small surface, heavier nuclei are close to nuclear matter and
hence are ideal to study the physics.

The basic principle behind this algorithm \cite{saca} is that
fragment structure was achieved via energy minimization using
\emph{simulated annealing technique} which yields maximum binding
energy of the system consisting of fragments of all sizes produced
in a reaction. In this algorithm, each cluster is subjected to a
binding energy check. As a first attempt, a constant average
binding energy check of -4 MeV/nucleon was employed for the all
clusters. This algorithm (labeled as \emph{simulated annealing
clusterization algorithm} i.e. SACA) yielded quite encouraging
results. For instance, one could explain the fragment distribution
for the reactions of O+ Ag/Br at incident energies 25-200 AMeV
\cite{jai,jai2}. For the first time, this microscopic approach
\cite{gosx} could also reproduce the fragment multiplicities in
$^{197}Au+ ^{197}Au $ reaction at E=600 AMeV measured by ALADiN
collaboration. It is worth mentioning that the MST approach failed
badly to reproduce this experimental trend \cite
{gosx,blaich,tsang}. A comparison of SACA (without binding energy
cut of -4 MeV/nucleon) and one developed by Dorso \emph{et al}.
\cite{dorso} yielded the same results for lighter colliding
nuclei.

As discussed above, each fragment in SACA method was subjected to
a constant binding energy of -4 MeV/nucleon. We know that the
binding energy depends on the mass of the fragment/nucleus. One is
always wondering whether this criterion of average binding energy
is justified or not. In this paper, we wish to address the above
question by subjecting each fragment to its true binding energy
that has now been measured to a very precise level with reference
to unstable and stable isobars, proton-rich and neutron-rich
nuclei. We shall show that this improvement does not yield
different results justifying the validity of the algorithm.

We employ quantum molecular dynamics (QMD) model as primary model
to follow the time evolution of nucleons. Section \ref{model}
describes the primary QMD model along with details of {\it
{simulated annealing clusterization algorithm}} (SACA) and its
extension. Section \ref{result} deals with the calculations and
illustrative results, which are summarized in section
\ref{summary}.

\section{\label{model} The Model}
\subsection{Quantum Molecular Dynamics (QMD) model}

The quantum molecular dynamics is an \emph{n-}body theory that
simulates the heavy-ion reactions between 30 AMeV and 1AGeV on
event by event basis. This is based on a molecular dynamics
picture where nucleons interact via two and three-body
interactions. The explicit two and three-body interactions
preserve the fluctuations and correlations which are important for
\emph{n}-body phenomenon such as multifragmentation
\cite{aich,rkp,hart}. Nucleons follow the classical trajectories
obtained by Hamilton's equations of motion:
\begin{eqnarray}
\dot{{\bf r}_\alpha}&=&{\bf {\nabla}_p}_{\alpha}\langle {\cal H}
\rangle, \alpha=1,. . .
,N; \nonumber \\
\dot{{\bf p}_\alpha}&=&-{\bf {\nabla}_r}_{\alpha}\langle {\cal H}
\rangle, \alpha=1,. . . ,N. \label{euler}
\end{eqnarray}
Here, nucleons interact via \emph{n-n} interactions and stochastic
elastic and inelastic collisions. For further details of the
model, the reader is referred to Ref. \cite{aich}.

\subsection{Extended SACA formalism}
As discussed in the previous section, earlier versions of
clustering algorithm such as minimum spanning tree (MST) rely on
the spatial correlation principle to identify the fragment
configuration \cite{aich}. In this algorithm, two nucleons are
considered to be a part of the same fragment if their
inter-nucleon distance is smaller than $r_{C}$ (in fm) . One
generally takes $2\leq r_{C} \leq 4$. Naturally, it cannot address
the time scale of fragmentation. This failure led to the
development of more sophisticated algorithm based on the simulated
annealing technique. This approach, known as \emph{simulated
annealing clusterization algorithm} (SACA), is based on the
principle of energy minimization which requires that a group of
nucleons can form a bound fragment if their total fragment energy
per nucleon $\zeta _{i}$ is below certain binding energy
$E_{bind}$ \emph{i.e.}
\begin{eqnarray}
\zeta_{i}=\frac{1}{N_{f}}\sum_{\alpha=1}^{N_{f}}
\left[\sqrt{\left(\textbf{p}_{\alpha}-\textbf{P}_{N_{f}}
\right)^{2}+m_{\alpha}^{2}}-m_{\alpha} + \frac{1}{2}\sum_{\beta
\neq \alpha}^{N_{f}}V_{\alpha \beta}
\left(\textbf{r}_{\alpha},\textbf{r}_{\beta}\right)\right]<-E_{Bind}.
\label{be}
\end{eqnarray}
In the original SACA version \cite{saca}, we take $E_{bind}$ = 4.0
AMeV if $N_{f}\geq3$ and $E_{bind} = 0$ otherwise. In this
equation, $N_{f}$ is the number of nucleons in a fragment,
$P_{N_{f}}$ is the average momentum of the nucleons bound in the
fragment. To find the most bound configuration, we start with a
random configuration and the energy of each cluster is calculated
using Eq. (\ref{be}). Let the total energy of a configuration {\it
k} be $E_{k}$ (=$ {\sum_{i}}N_{f}\zeta_{i}$), with $\zeta_{i}$ is
the energy per nucleon associated with that fragment.

Now to generate new configuration $k^{'}$, we assume that this can
be achieved by (a) transferring a nucleon from some randomly
chosen fragment to another fragment, by (b) setting a nucleon free
or, by (c) absorbing a free nucleon into a fragment] has total
energy $E_{k}^{'}$. If the difference between energies of the old
and the new configurations, $\Delta E (= E_{k}^{'}-E_{k})$ is
negative, the new configuration is always accepted. If not, the
new configuration $k^{'}$ may nevertheless be accepted with a
probability of $exp~(-\Delta E/c)$, where `c' is called control
parameter. This procedure is known as Metropolis algorithm. The
control parameter is decreased in small steps. This algorithm will
yield eventually the most bound configuration (MBC). Since this
combination of Metropolis algorithm with decreasing control
parameter is known as simulated annealing, this approach was
dubbed as simulated annealing clusterization algorithm (SACA)
\cite{saca}. The present algorithm with a constant average binding
energy check is labeled as SACA (1.1). For further details, we
refer the reader to Refs. \cite{saca,jai,jai2,saca1}.
\begin{figure}[!t]
\begin{center}
\vskip -0.8 cm
\includegraphics*[scale=0.5] {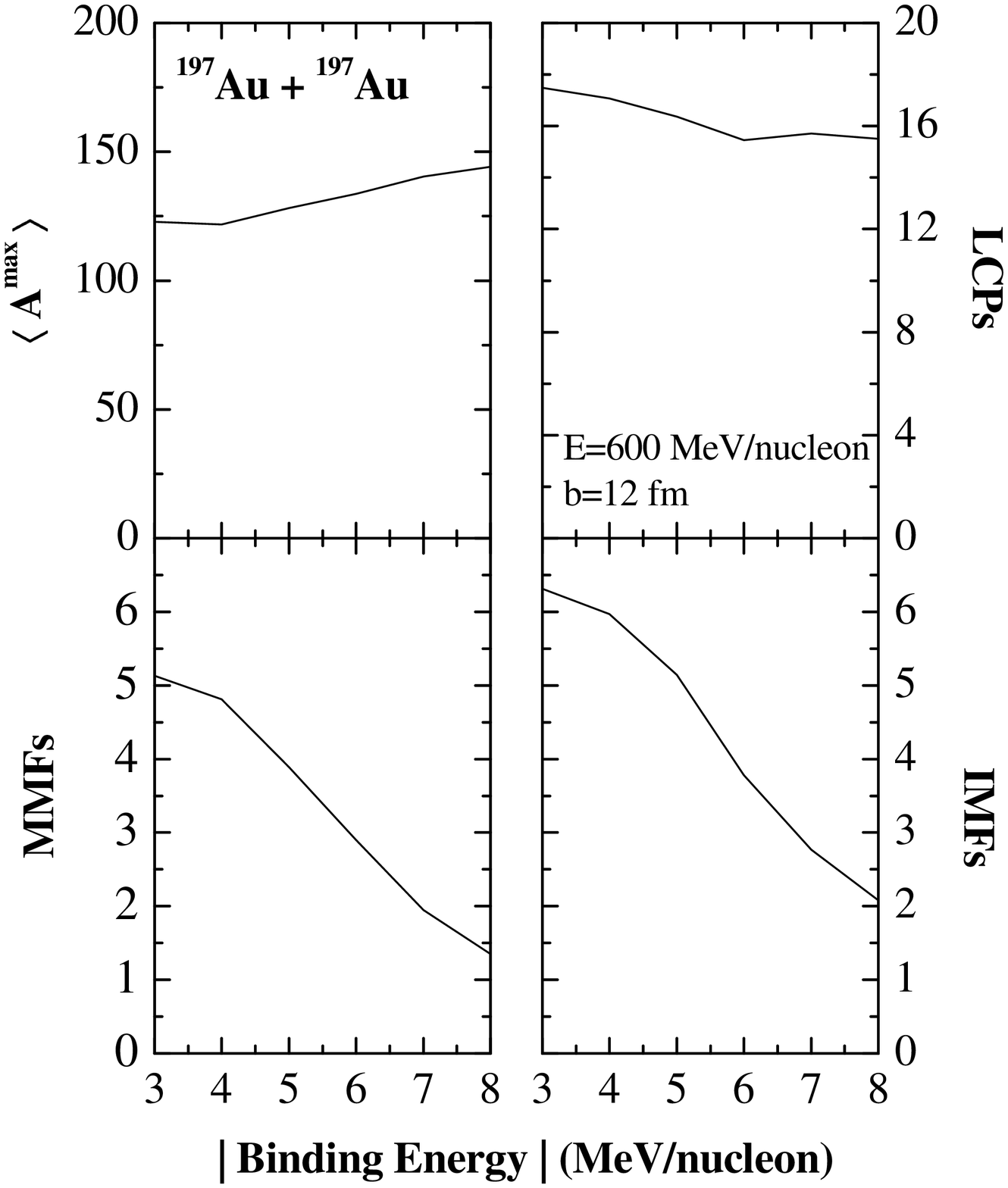}
\vskip -0.8 cm \caption {\label{bec}The average mass of the
heaviest fragment $\langle A^{max} \rangle $, mean multiplicities
of light charged particles LCPs, medium mass fragments MMFs, and
intermediate mass fragments IMFs as a function of binding energy
check imposed for the reaction of $^{197}Au+^{197}Au$ at 600
MeV/nucleon and at an impact parameter of 12 fm. }
\end{center}
\vskip -0.4 cm
\end{figure}
In Fig.~\ref{bec}, we show the calculated multiplicities of
different fragments as well as the mean size of the largest
fragment $\langle A^{max} \rangle$ as a function of different
binding energy cuts using original SACA. The multiplicities of
light charged particles LCPs $[2\leq A \leq 4]$ and size of the
largest fragment $\langle A^{max} \rangle$ remains almost
unaffected by changing the binding energy cut. However, the
multiplicities of medium mass fragments MMFs $[5\leq A \leq 20]$
and intermediate mass fragments IMFs $[5\leq A \leq 65]$ show
strong sensitivity towards the imposed binding energy. From this
analysis, it would be interesting to study the effect of binding
energy on average fragment production. The choice of proper
binding energy can be based on either experimental information or
on theoretical information. Since experimental information is
range bound, we shall use theoretical formulation.

One of the earlier attempts to reproduce the gross features of
nuclear binding energies was made by Weizs\"{a}cker {\it et al.}
\cite{weiz}. The Bethe-Weizs\"{a}cker {\it (BW)} mass formula for
the binding energy of a nucleus reads as \cite{sam}:
\begin{eqnarray}
E_{bind} = &a_{v}N_f - a_{s}N_{f}^{2/3} - a_{c}
\frac{N_{f}^{z}(N_{f}^{z}-1)}{N_{f}^{1/3}} - a_{sym}\frac{(N_f
-2N_{f}^{z})^{2}}{N_{f}} + \delta.
\end{eqnarray}
Here, $N_{f}^{z}$ stands for the proton number of a fragment. The
various terms involved in this mass formula are the volume,
surface, Coulomb, asymmetry and pairing terms. The strength of
different parameters is: $a_{v}$=15.777 MeV, $a_{s}$=18.34 MeV,
$a_{c}$=0.71 MeV and $a_{sym}$=23.21 MeV respectively \cite{sam}.
The pairing term $\delta$ is given by:
\begin{equation}
\delta=+a_{p}N_{f}^{-1/2}~ for~even~N_{f}^{z}~and~even~N_{f}^{n}, \\
\end{equation}
\begin{equation}
\delta=-a_{p}N_{f}^{-1/2}~ for~odd~N_{f}^{z}~and~odd~N_{f}^{n},\\
\end{equation}
\begin{equation}
\delta=0~ for~ odd~N_{f}~nuclei,
\end{equation}
with $a_{p}$ = 12 MeV and $N_{f}^{n}$ being the neutron number of
a fragment. This formula reproduces the binding energy of stable
nuclei but faces serious problem for light nuclei along the drip
line and with nuclei having rich neutron or proton content. The
inadequacy of BW mass formula for lighter nuclei was removed by
Samanta {\it et al.} \cite{sam} by modifying its asymmetry and
pairing terms. This modified formula was dubbed as modified
Bethe-Weizs\"{a}cker mass (BWM) formula \cite{sam}. The beauty of
BWM formula lies in its ability to reproduce the binding energies
for light nuclei near the drip line \cite{sam}. For a large number
of unstable isobars, isotones and halo nuclei, it was shown in
Ref. \cite{sam} that this modified formula reproduces the
experimental binding energies quite precisely. In the BWM formula,
the binding energy of a fragment is defined as \cite{sam}:
\begin{eqnarray}
 E_{bind}=& a_{v}N_{f} - a_{s}N_{f}^{2/3} -
a_{c}\frac{N_{f}^{z}(N_{f}^{z}-1)}{N_{f}^{1/3}}
-a_{sym}\frac{(N_{f}-2N_{f}^{z})^{2}}{N_{f}(1+e^{-N_{f}/17})} +
\delta_{new}. \label{bwn}
\end{eqnarray}
The strength of various parameters now reads: $a_{v}$=15.777 MeV,
$a_{s}$=18.34 MeV, $a_{c}$=0.71 MeV and $a_{sym}$=23.21 MeV,
respectively. The pairing term $\delta_{new}$ is given by:
\begin{equation}
\delta_{new}=+a_{p}N_{f}^{-1/2}~(1-e^{-N_{f}/30})~for~even~N_{f}^{z}~and~even~N_{f}^{n},\\
\end{equation}
\begin{equation}
\delta_{new}=-a_{p}N_{f}^{-1/2}(1-e^{-N_{f}/30})~for~odd~N_{f}^{z}~and~odd~N_{f}^{n},\\
\end{equation}
\begin{equation}
\delta_{new}=0~ for~odd ~N_{f}~nuclei,
\end{equation}
with $a_{p}$ = 12 MeV.

We extend the SACA method by incorporating this binding energy
formula during the formation of the clusters. Each fragment at the
end of the procedure is subjected to this new binding energy
(Eq.(\ref{bwn})) instead of a constant -4 MeV/nucleon binding
energy. Any fragment that fails to fulfil the above binding energy
criterion is treated as a group of free nucleons. At the end, all
fragments are properly bound. This version is labeled as SACA
(2.1). We have also tested the spectrum for actual experimental
binding energies \cite{audi95}. Only small difference is seen for
lighter fragments only.

\section{\label{result}Results and Discussions}

\begin{figure}
\begin{center}
\includegraphics [scale=0.5] {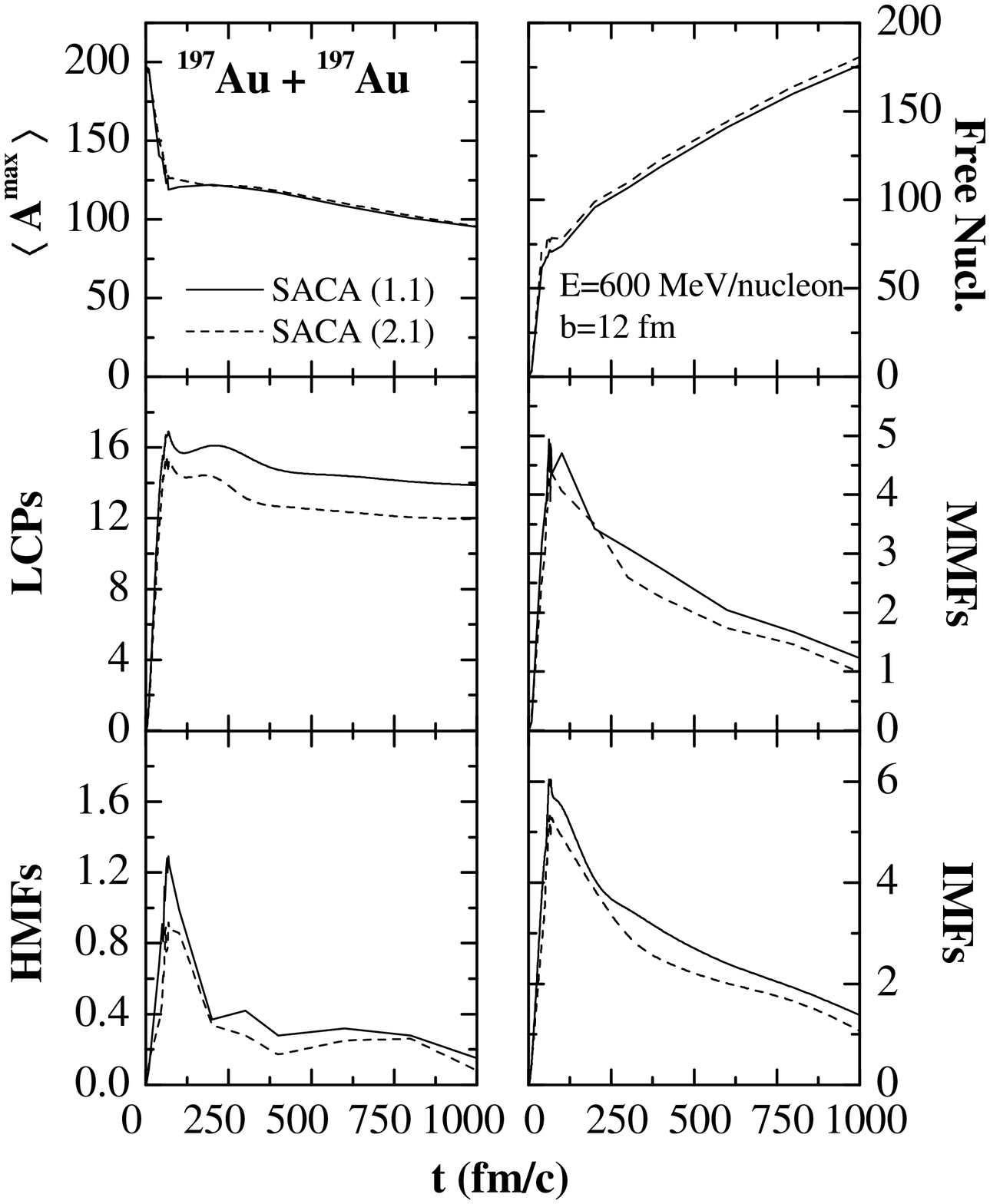}
\vskip -1.0cm \caption {\label{t1} The average mass of heaviest
fragment $\langle A^{max} \rangle $ and the mean multiplicities of
various kinds of fragments as a function of time for the reaction
of $^{197}Au +^{197}Au$ at 600 MeV/nucleon and at an impact
parameter of 12 fm. The solid and dashed lines depict the results
due to original SACA and its extension.}
\end{center}
\end{figure}

We simulated the collisions of $^{197}Au+ ^{197}Au$ at incident
energy of 600 AMeV using a \emph{soft} equation of state along
with standard energy dependent \emph {n-n} cross section. We
display in Fig.~\ref{t1}, the average mass of the largest fragment
$\langle A^{max} \rangle $, mean multiplicities of free nucleons,
light charged particles LCPs $[2\leq A \leq 4]$, medium mass
fragments MMFs $[5\leq A \leq 20]$, heavy mass fragments HMFs $[21
\leq A \leq 65]$ and intermediate mass fragments IMFs $[5 \leq A
\leq 65]$ as a function of time for the reaction of $^{197}Au+
^{197}Au $ at 600 AMeV and impact parameter 12 fm. As expected,
$\langle A^{max} \rangle$ is nearly independent of the binding
energy criterion, whereas insignificant influence can be seen on
the multiplicities of free nucleons, LCPs, MMFs and IMFs. Similar
trends were also observed for the central reaction of $^{197}Au+
^{197}Au $ at 600 AMeV.

\begin{figure}
\begin{center}
\includegraphics [scale=0.5] {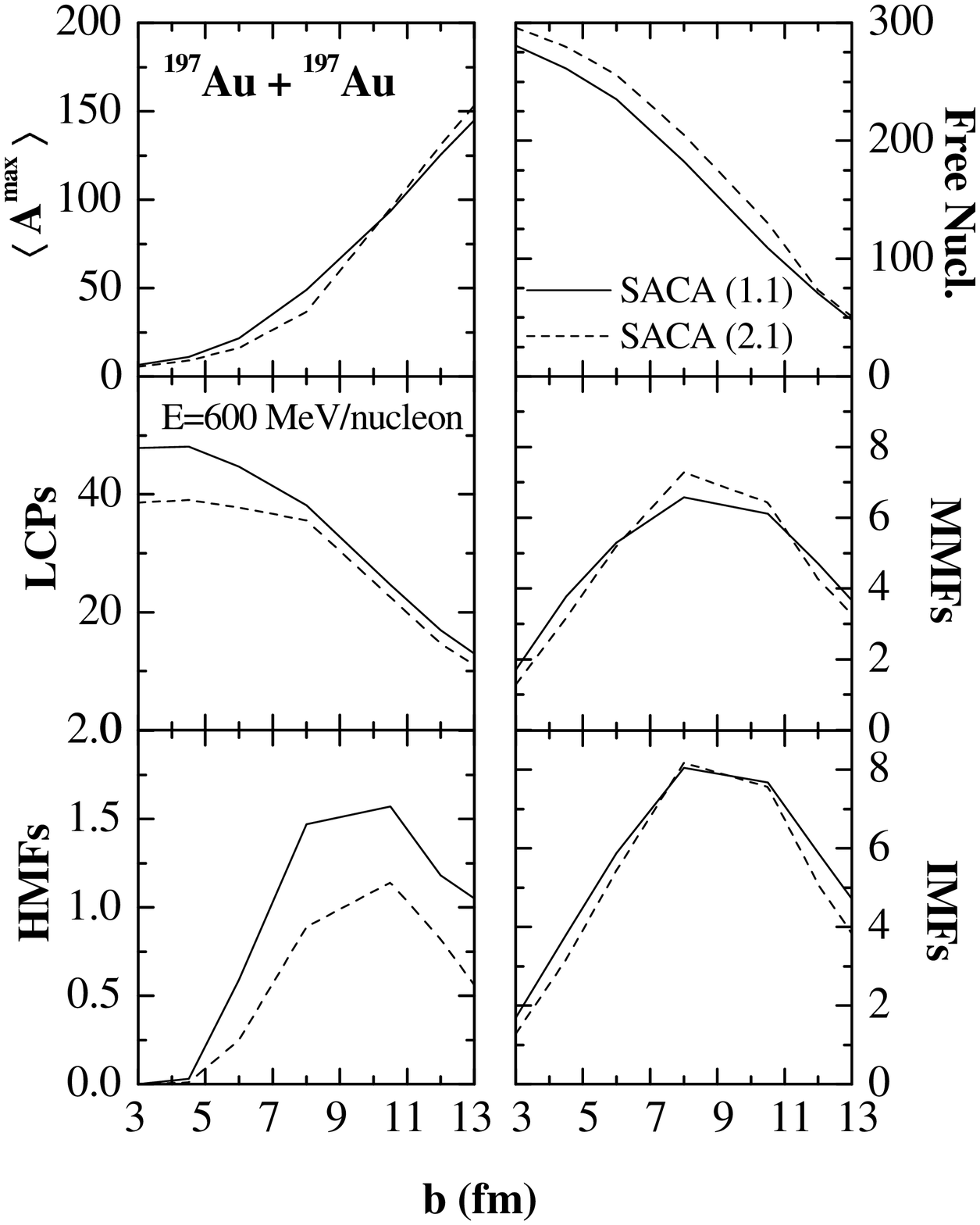}
\vskip -0.6 cm \caption {\label{impact} The impact parameter
dependence of average size of the heaviest fragment $\langle
A^{max} \rangle $ and mean multiplicities of various kinds of
fragments for the reaction of $^{197}Au+^{197}Au$ at incident
energy 600 MeV/nucelon. The solid and dashed curves depict results
of SACA (1.1) and SACA (2.1), respectively.}
\end{center}
\end{figure}

\begin{figure} [!t]
\begin{center}
\includegraphics [scale=0.5] {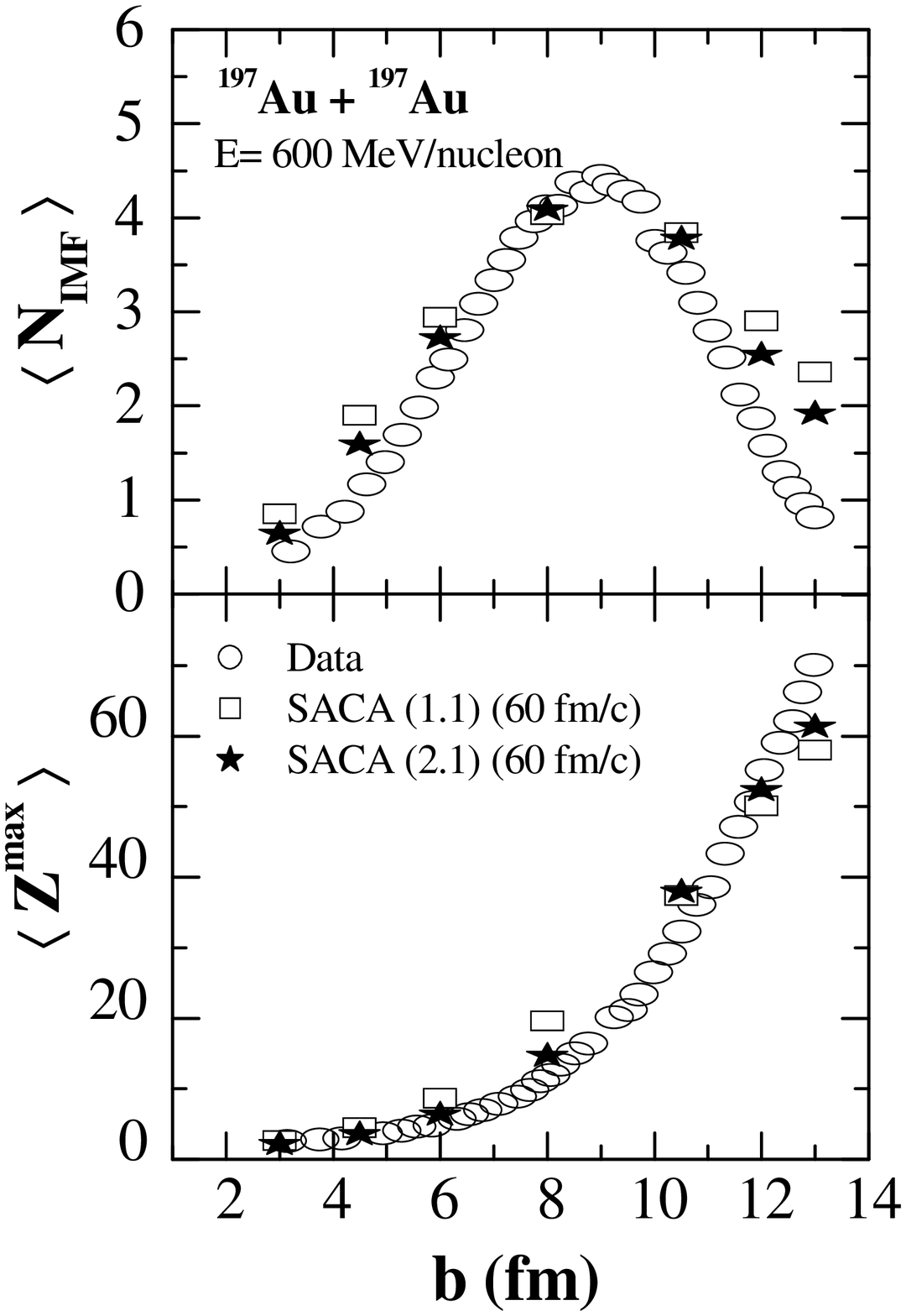}
\vskip -1.2cm \caption {The mean IMF multiplicity (top panel) and
average charge of the heaviest fragment (bottom panel) as a
function of impact parameter. Open circles depict the experimental
data points \cite{schut}. \label{imf} }
\end{center}
\end{figure}
To further explore the characteristics of fragment structure
obtained with modified SACA (2.1), we show in Fig.~\ref{impact},
the impact parameter dependence of mean multiplicities of various
fragments. This will also help to understand the proper energy
deposition in the spectator matter. The result obtained with SACA
(1.1) and SACA (2.1) are displayed for the reaction of $^{197}Au+
^{197}Au$ at 600 AMeV as a function of impact parameter. The time
for realization of different fragments was chosen to be 60 fm/c.
This is the time when $\langle A^{max} \rangle $ has minimum size
and configuration realized at this stage is most bound
\cite{saca}. In central collisions, SACA (2.1) predicts smaller
$\langle A^{max} \rangle $, whereas trend reverses in the
peripheral collisions. As a result, free nucleons also behave
accordingly. The yields of IMFs and MMFs do not reduce appreciably
for central as well as peripheral geometries using extended
version of SACA. This is due to the fact that fragments recognized
by SACA method are properly bound, therefore, simple cut also
yields same results.

We also attempted to confront our present calculations using
extended clusterization approach SACA (2.1) (at t=60 fm/c) with
experimental data of ALADiN group \cite{schut} for the reaction of
Au (600 AMeV) + Au. In Fig.~\ref{imf}, we show the mean IMF
multiplicity $\langle N_{IMF} \rangle $ (in upper panel) and
average charge of the largest fragment $\langle Z^{max} \rangle$
(in lower panel) as a function of impact parameter at 600 AMeV.
The calculations with the original SACA (1.1) version are also
shown for comparison. All calculations were subjected to
experimental cuts of forward hemisphere. The $\langle N_{IMF}
\rangle $ and $\langle Z^{max} \rangle$ obtained with different
versions of SACA are quite close to each other and to the
experimental data. It justifies the use of average binding energy
within above algorithm. We have also calculated the yields at
incident energies of 400 and 1000 MeV/nucleon. Similar results are
also obtained at these incident energies.

\section{\label{summary}Summary}
Summarizing the work, we have proposed an extension to SACA method
by incorporating the binding energy of individual fragments
calculated from the modified Bethe-Weizs\"{a}cker mass (BWM)
formula. Based on our calculations, we noticed that this extension
has little effect on the fragment multiplicities and mean size of
the largest fragment at 60 fm/c as well as at asymptotic times. In
peripheral collisions, new extension reduces the IMF yield,
thereby increasing the size of $\langle A^{max} \rangle $
marginally. Both versions of SACA are clearly close to each other
and to ALADiN data. \\

One of the authors (Y. K. V) would like to thank Drs. W. Trautmann
and C. Samanta for interesting and constructive discussions. This
work is supported by a research grant from Council of Scientific
and Industrial Research, Government of India, vide grant no.
7167/NS-EMR-II/2006.

\end{document}